\documentclass[11pt,pra,aps,onecolumn,superscriptaddress,floatfix,notitlepage,nofootinbib,longbibliography]{revtex4-1}
\usepackage[latin1]{inputenc}
\usepackage{mathrsfs}
\usepackage{amsmath,dsfont}
\usepackage{amsfonts,natbib}
\usepackage{amssymb}
\usepackage{bm}
\usepackage[pdftex]{graphicx}
\usepackage{comment}

\usepackage[colorlinks=true,allcolors=blue]{hyperref}
\begin{document}

\title{Detector Correlations and Null Tests of the Coherent State Hypothesis}
\author{Sreenath K. Manikandan}
\email{sreenath.k.manikandan@su.se}
\affiliation{Nordita, Stockholm University and KTH Royal Institute of Technology, Hannes Alfv\'{e}ns v\"{a}g 12, SE-106 91 Stockholm, Sweden}
\author{Frank Wilczek}
\email{fwilczek@asu.edu}
\affiliation{Department of Physics, Arizona State University, Tempe, Arizona 25287, USA}
\affiliation{T. D. Lee Institute, Shanghai 201210, China}
\affiliation{Wilczek Quantum Center and Department of Physics and Astronomy, Shanghai Jiao Tong University, Shanghai 200240, China }
\affiliation{Department of Physics, Stockholm University, AlbaNova University Center, 106 91 Stockholm, Sweden}
\affiliation{Nordita, Stockholm University and KTH Royal Institute of Technology, Hannes Alfv\'{e}ns v\"{a}g 12, SE-106 91 Stockholm, Sweden}
\affiliation{Center for Theoretical Physics, Massachusetts Institute of Technology, Cambridge, Massachusetts 02139, USA}
\date{\today}

\begin{abstract}
    We discuss the statistics of correlations between two resonant detectors.  We show that this allows simple null tests of the coherent state hypothesis, free of vacuum (quantum) noise.  Complementary aspects of the radiation field, {\it e.g.}, squeezing in number or phase, can be revealed through appropriate detection strategies.  
\end{abstract}

\maketitle
\section{Introduction}

Coherent states are the default option for modeling ``classical'' radiation fields in quantum theory. As left and right eigenstates, respectively, of the field creation and annihilation operators, coherent states allow us to replace field expectation values by c-numbers, thus enabling a seamless transition between semiclassical and quantum descriptions of radiation fields and their interactions with matter. This correspondence is formalized in the optical equivalence theorem of quantum optics~\cite{Sudharshan,Glauber}. Conversely, observable phenomena that indicate violation of the coherent state hypothesis indicate the need to abandon a purely classical description.  As we have recently emphasized, such phenomena can occur even  at large occupation numbers of the mode under study.   These issues are especially salient for gravitational radiation~\cite{Manikandan_Wilczek_acoherence,ParikhPRD,ParikhPRL,ParikhEPJD,guerreiro_quantum_2020,Guerreiro2022quantumsignaturesin,guerreiro_nonlinearities_2023,Essay,tobar_detecting_2024,tobar_detecting_2025,loughlin_wave-particle_2025,CarneyPRD,carney_comments_2024}.

%%%%%%%%%%

We will be using resonant mass detectors to design statistical tests for the acoherence of gravitational radiation. They have been suggested recently as a viable strategy to observe quantized response to gravitational radiation~\cite{tobar_detecting_2024,shenderov_stimulated_2024}. The response of such detectors is probabilistic, and we have demonstrated that the probability distribution gives us access to basic information about the quantum nature of the radiation field ~\cite{Manikandan_Wilczek_acoherence,Essay,manikandan_complementary_2025}. 

Because the length scale of the acoustic resonances is governed by are the speed of sound ($v_s$) as opposed to the speed of light (c) it possible for an array of several detectors to fit well within the one wavelength of the resonant gravitational radiation. Here we show that cross-correlations between initially uncorrelated resonant mass detectors provide additional insight.  Notably, we find that this provides null tests for the coherent state hypothesis.  On the other hand, non-zero correlations of different kinds allow us to probe several aspects of the radiation field's quantum state.  

Whereas in our earlier work, violation of the coherent state hypothesis showed up as excess noise~\cite{ParikhEPJD,ParikhPRL,ParikhPRD,manikandan_complementary_2025,Manikandan_Wilczek_acoherence,Essay}, with non-zero vacuum noise as a background, here we encounter no such background.

Our calculations belong to the circle of ideas around the Hanbury-Brown-Twiss experiment~\cite{brown_lxxiv_1954,brown_correlation_1956,hanbury_brown_test_1956}, where we consider a simpler setup using a single source and two detectors interacting with a single effective mode of the gravitational radiation field.  We show that the cross correlation between two detectors allows us to estimate the second-order coherence function $g^{(2)}$ for the gravitational radiation field. 

Although we focus on cross-correlations between resonant mass detectors for gravitational radiation, one can envision similar tests using interferometric methods~\cite{Maulikcross}.

\section{Correlation of counts}
We extend the simple model detailed in Ref.~\cite{Manikandan_Wilczek_acoherence} to the case of two parallel detectors (labeled by annihilation operators $d_{i},~i=1,2$), and describe their interaction with a common, effectively single mode of the radiation field, labeled by the annihilation operator, $a$, to which the detectors are optimally aligned. The interaction between each detector and the radiation field assumes the form,
\begin{equation}
    H_{I,1} \Delta t_1 = \hbar\sqrt{\gamma_0\Delta t_1}(a^\dagger b_1+b_1^\dagger a),
\end{equation}
and,
\begin{equation}
    H_{I,2} \Delta t_2 = \hbar\sqrt{\gamma_0\Delta t_2}(a^\dagger b_2+b_2^\dagger a).
\end{equation}
Above $\gamma_0 = \frac{8GML^2\omega^4}{\pi^4 c^5}$ is the spontaneous emission rate for gravitons for the fundamental acoustic mode of resonant mass detectors having frequncy $\omega$ (here $G:$ Newton's constant, $M:$ the mass, and $L:$ the length of the resonant mass detector, and $c:$ the speed of sound). The spontaneous emission rate $\gamma_0$ is taken to be the same for both detectors, assuming identical detectors. We can consider that both detectors are initialized in vacuum and the radiation field is initially in a generic quantum state represented by the diagonal $P$-representation in the coherent state basis~\cite{Sudharshan,Glauber},
\begin{equation}
    \rho_{F}=\int d^{2}\alpha
P(\alpha)|\alpha\rangle\langle\alpha|,~~\rho = \rho_{F}\otimes|0\rangle\langle 0|_{1}\otimes |0\rangle\langle 0|_{2}.\end{equation}
The time-evolved state after interaction with the detectors can be approximated as (see Appendix.~\ref{AppA}),
\begin{equation}
    \rho\rightarrow\rho'\approx \int d^{2}\alpha P(\alpha)|\alpha\rangle\langle\alpha|\otimes |-i\alpha\sqrt{\gamma_{0}\Delta t}\rangle\langle -i\alpha\sqrt{\gamma_{0}\Delta t}|_1\otimes |-i\alpha\sqrt{\gamma_{0}\Delta t}\rangle\langle -i\alpha\sqrt{\gamma_{0}\Delta t}|_2.\label{state2}
\end{equation}
The approximation is valid for $\gamma_0\Delta t\ll 1$ (which turns out to be the case for resonant mass detectors for gravitational radiation~\cite{Manikandan_Wilczek_acoherence,tobar_detecting_2024}) and complete results are provided in Appendix.~\ref {AppA}. From this, we can compute the relevant quantities for estimating the detector cross-correlations between two click detectors as our first example. We need to estimate the following quantities,
\begin{equation}
  \langle N_1\rangle =   \langle b_1^\dagger b_1\rangle, ~~\langle N_2\rangle =   \langle b_2^\dagger b_2\rangle,~~\&~~\langle N_1 N_2\rangle =  \langle b_1^\dagger b_1b_2^\dagger b_2\rangle.
\end{equation}
The joint probability distribution of clicks is given by,
\begin{eqnarray}
    p(n_1,n_2)&=& \text{tr}_F\langle n_1,n_2|\rho'|n_1,n_2\rangle \nonumber\\
    &=&\frac{1}{n_1!n_2!}\int d^2\alpha P(\alpha)[\gamma_0\Delta t|\alpha|^2]^{n_1}e^{-|\alpha|^2\gamma_0\Delta t} [\gamma_0\Delta t|\alpha|^2]^{n_2}e^{-|\alpha|^2\gamma_0\Delta t}.
\end{eqnarray}

We find that,
\begin{equation}
    \langle N_i\rangle = \sum_{n_1,n_2}p(n_1,n_2)n_i \approx (\gamma_0\Delta t)\langle a^\dagger a\rangle, ~~i=1,2.
\end{equation}
We also obtain,
\begin{eqnarray}
  \langle N_1 N_2\rangle &\approx&\sum_{n_1,n_2}p(n_1,n_2)n_1n_2\nonumber\\&=&(\sqrt{\gamma_0\Delta t})^4\int d^2\alpha P(\alpha)|\alpha|^4=   (\sqrt{\gamma_0\Delta t})^4\langle (a^\dagger)^2 a^2\rangle\nonumber\\
  &=& (\gamma_0\Delta t)^2 (Q\langle a^\dagger a\rangle +\langle a^\dagger a\rangle^2), 
\end{eqnarray}
where the Mandel's $Q$ parameter is defined as, $Q=\left[\langle (a^\dagger a)^2\rangle-\langle a^\dagger a\rangle^2-\langle a^\dagger a\rangle\right]/\langle a^\dagger a\rangle$. We are interested in correlator of counts, which can now be estimated as,
\begin{equation}
    \langle N_1 N_2\rangle -\langle N_1\rangle \langle N_2\rangle  = (\gamma_0\Delta t)^2Q\langle a^\dagger a\rangle.
\end{equation}
We observe that the correlator of counts is a direct probe for the quantum nature of the radiation field; the correlator vanishes for coherent states ($Q=0$), providing the desired null outcome of interest for coherent states. Practically, however, sub-Poissonian states of gravitational radiation (having $-1\leq Q <0$) will also show negligible correlation. This is because the graviton to click conversion efficiency for resonant mass detectors is very small, $\gamma_0\Delta t \ll 1$, reflecting the smallness of the spontaneous emission rate~\cite{Manikandan_Wilczek_acoherence,manikandan_complementary_2025,Essay,tobar_detecting_2024}. Nevertheless clicks (stimulated absorption) can be observed with high probability when $\gamma_0\Delta t\langle a^\dagger a\rangle \sim O(1)$; the large number of gravitons $\langle a^\dagger a\rangle\gg 1$ required to compensate for the smallness of detector efficiency can be supplied by known sources of gravitational radiation in the LIGO band  (kilo-Hertz frequencies, and amplitude $h_0\sim 10^{-22}$)~\cite{tobar_detecting_2024,shenderov_stimulated_2024}. 

Thus super-Poissonian states having $Q\sim \langle a^\dagger a\rangle$ can show measurable deviations in the null-test using click detectors, even -- indeed, especially -- in the limit of large number of occupancy for the radiation mode,  $\langle a^\dagger a\rangle$. 

The second-order coherence function for the gravitational radiation field appears directly in the ratio,
\begin{equation}
   R=\frac{ \langle N_1 N_2\rangle}{\langle N_1\rangle\langle N_2\rangle} = \frac{\langle (a^\dagger)^2 a^2\rangle}{\langle a^\dagger a\rangle^2} = 1+\frac{Q}{\langle a^\dagger a\rangle} = g^{(2)}(0).
\end{equation}
Since $g^{(2)}(0)$ can also be estimated using the lowest order click probabilities for resonant mass detectors via the relation~\cite{Manikandan_Wilczek_acoherence},
\begin{equation}
    R=\frac{2P_2P_0}{P_1^2}\approx g^{(2)}(0),
\end{equation}
estimating $g^{(2)}(0)$ using detector cross-correlations affords a consistency check and (given consistency) a way to decrease sampling error. 

We can also define a first-order coherence function associated with the state $\rho'$ in Eq.~\eqref{state2} as, $|g^{(1)}(0)|=|\langle b_1^\dagger b_2\rangle|/\sqrt{\langle b_1^\dagger b_1\rangle\langle b_2^\dagger b_2\rangle}=1$, and this holds independent of the state of the radiation field $\rho_F$ (it is simply unity for any state of the radiation field as we assume a single mode of the radiation field, and no space/time delay between the two detectors). The above click measurement strategy for each detector alone is insufficient to measure $g^{(1)}(0)$; however can be achieved, for example, by changing experimental contexts to also measure clicks in the collective sum and difference modes of the two detectors.
\section{Correlations in the Homodyne signals}
We now consider the cross-correlations between the readouts of two resonant mass detectors operating as Homodyne detectors. We begin with the same scenario as before, such that the joint state of the radiation field and two detectors after two sequential interactions is given by $ \rho'$ in Eq.~\eqref{state2}.

We can use this to obtain the joint probability distribution $p(x_1,x_2)$ for Homodyne measurement quadratures, $x_1,x_2$. We obtain that
\begin{eqnarray}
    p(x_1,x_2)&=& \text{tr}_F\langle x_1,x_2|\rho'|x_1,x_2\rangle \nonumber\\
    &\approx&\frac{1}{\pi x_{0_1}x_{0_2}}\int d^2\alpha P(\alpha) e^{-\frac{[x_1-\sqrt{2}x_{0_1} \text{Im}(\alpha)\sqrt{\gamma_0\Delta t}]^2}{x_{0_1}^2}}e^{-\frac{[x_2-\sqrt{2}x_{0_2} \text{Im}(\alpha)\sqrt{\gamma_0\Delta t}]^2}{x_{0_2}^2}},
\end{eqnarray}
where $x_{0_i}\propto\sqrt{\hbar},~i=1,2$ corresponds to the zero point length of each of the detectors. Using the above, we also obtain,
\begin{eqnarray}
    \langle x_i\rangle  &\approx& \int dx_1 dx_2 p(x_1,x_2)x_i =  \sqrt{2}x_{0_i}\sqrt{\gamma_0\Delta t} \int d^2\alpha  P(\alpha)\text{Im}(\alpha)\nonumber\\ &=& \sqrt{2}x_{0_i}\sqrt{\gamma_0\Delta t}\langle \text{Im}(\alpha)\rangle,~~i=1,2,
\end{eqnarray}
and,
\begin{eqnarray}
    \langle x_1 x_2\rangle  &=& \int dx_1 dx_2 p(x_1,x_2)x_1 x_2 =  2x_{0_1}x_{0_2}\gamma_0\Delta t \int d^2\alpha  P(\alpha)\text{Im}(\alpha)^2\nonumber\\ &=&2x_{0_1}x_{0_2}\gamma_0\Delta t  \langle\text{Im}(\alpha)^2\rangle.
\end{eqnarray}
We now see that the detector cross-correlation function,
\begin{eqnarray}
    \langle x_1 x_2\rangle -\langle x_1\rangle\langle x_2\rangle  &=& 2x_{0_1}x_{0_2}\gamma_0\Delta t[\langle\text{Im}(\alpha)^2\rangle-\langle\text{Im}(\alpha)\rangle^2]\nonumber\\
    &=&x_{0_1}x_{0_2}\gamma_0\Delta t\bigg(\langle (\Delta \hat{P})^2\rangle -\frac{1}{2}\bigg),~~\hat{P}=\frac{a-a^\dagger}{\sqrt{2}i}.
\end{eqnarray}
The cross correlation vanishes if the radiation field is in a coherent state $\langle (\Delta \hat{P})^2\rangle=\frac{1}{2}$, providing the desired null outcome. Interestingly the correlator can be of order unity when the radiation field is in a number state, $\langle (\Delta \hat{P})^2\rangle=n+\frac{1}{2}$, provided the energy density (or the number of incoming gravitons) are substantially large to compensate for the small spontaneous emission rate $\gamma_0$ [i.e., when $n\gamma_0\Delta t\sim O(1)$]. This is possible if the gravitational radiation has energy densities comparable to those observed at LIGO~\cite{manikandan_complementary_2025,Manikandan_Wilczek_acoherence,tobar_detecting_2024,Essay,shenderov_stimulated_2024}. Importantly, while a null outcome for the correlator estimate between two click detectors can be inconclusive for a Fock state as discussed previously, our findings above suggest that joint homodyne measurements can reveal statistical features that are complementary to click detectors in the quantum mechanical sense and allow us to further discriminate sub-Poissonian states (such as Fock states) from coherent states. 
\section{Joint measurements using Heterodyne detectors}
We proceed with our analysis to consider two phase-preserving (Heterodyne) measurements in sequence. The joint probability distribution of readouts is given by,
\begin{eqnarray}
    p(\beta_1,\beta_2)&=& \frac{1}{\pi^2}\text{tr}_F\langle \beta_1,\beta_2|\rho'|\beta_1,\beta_2\rangle \nonumber\\
    &\approx&\frac{1}{\pi^2}\int d^2\alpha P(\alpha) e^{-|\beta_1 +i\alpha \sqrt{\gamma_0\Delta t}|^2}e^{-|\beta_2 +i\alpha \sqrt{\gamma_0\Delta t}|^2}.
\end{eqnarray}
We find that,
\begin{eqnarray}
      \langle \text{Re}(\beta_i)\rangle  &=& \int d^2\beta_1 d^2\beta_2 \text{Re}(\beta_i)p(\beta_1,\beta_2)\nonumber\\
    &=&\sqrt{\gamma_0\Delta t}\int d^2\alpha P(\alpha)\text{Im}(\alpha) = \sqrt{\gamma_0\Delta t}\langle \text{Im}(\alpha)\rangle,~~i=1,2.
\end{eqnarray}
We can also evaluate the correlator,
\begin{eqnarray}
      \langle \text{Re}(\beta_1)\text{Re}(\beta_2)\rangle  &=& \int d^2\beta_1 d^2\beta_2 \text{Re}(\beta_1)\text{Re}(\beta_2)p(\beta_1,\beta_2)\nonumber\\
    &=&\gamma_0\Delta t\int d^2\alpha P(\alpha)\text{Im}(\alpha)^2 = \gamma_0\Delta t\langle \text{Im}(\alpha)^2\rangle.
\end{eqnarray}
We see that the cross-correlation between the two detectors yields,
\begin{eqnarray}
   \langle \text{Re}(\beta_1)\text{Re}(\beta_2)\rangle -\langle \text{Re}(\beta_1)\rangle\langle\text{Re}(\beta_2)\rangle  &=&\gamma_0\Delta t[\langle \text{Im}(\alpha)^2\rangle-\langle \text{Im}(\alpha)\rangle^2]\nonumber\\&=&\frac{1}{2}\gamma_0\Delta t\bigg(\langle (\Delta \hat{P})^2\rangle -\frac{1}{2}\bigg).
\end{eqnarray}
The detector cross-correlation is zero if the field is a coherent state, $\langle (\Delta \hat{P})^2\rangle =\frac{1}{2}$, leading to the desired null outcome. A state different from a coherent state would demonstrate substantial variability, making it a useful null test of acoherence. Although we have only estimated one of the cross-correlations as an example, the analysis can be extended in a straightforward way to estimate other cross-correlations between the real and imaginary parts of the two  Heterodyne signals.
Using $
\langle\beta_1^*\beta_2\rangle =\gamma_{0}\Delta t\langle |\alpha|^2\rangle=\gamma_{0}\Delta t\langle a^\dagger a\rangle$, and $
\langle\beta_i^*\rangle =i\sqrt{\gamma_{0}\Delta t}\langle \alpha^*\rangle=i\sqrt{\gamma_{0}\Delta t}\langle a^\dagger\rangle$, and $
\langle\beta_i\rangle =-i\sqrt{\gamma_{0}\Delta t}\langle \alpha\rangle=-i\sqrt{\gamma_{0}\Delta t}\langle a\rangle$, we can also estimate additional cross-correlators of the type,
\begin{equation}
\langle\beta_1^*\beta_2\rangle -\langle \beta_{1}^*\rangle\langle\beta_2\rangle=\gamma_{0}\Delta t\left[\langle |\alpha|^2\rangle -\langle\alpha^*\rangle\langle\alpha\rangle \right]=\gamma_{0}\Delta t\left[\langle a^\dagger a\rangle -\langle a^\dagger\rangle\langle a\rangle \right].
\end{equation}
The above correlator yields the null outcome, zero, for coherent states. It also captures the complementary features observable in the average signal; quantum mechanical states, such as Fock states or thermal states, for which the heterodyne signal vanishes on average, yield substantial deviation from zero in the above null test.

We also note that, unlike the quantum noise observable in a single heterodyne detector, where the measurement process itself can add excess quantum noise over the vacuum noise of the detector~\cite{arthurs_simultaneous_1965,loudon_squeezed_1987,clerk_introduction_2010,caves_quantum_1982,caves_quantum_2012,bergeal_phase-preserving_2010,haus_quantum_1962,manikandanFuel,manikandan_complementary_2025}, these cross correlations do not include the quantum noise from heterodyne detection or the detectors' vacuum noise, which is an added benefit.

\section{Conclusions}
We have demonstrated that independent detector cross-correlations can directly probe the quantum structure of gravitational radiation through null tests, where coherent states yield zero. The required parameter values for resonant mass detectors are such that our tests could be feasible. Specifically, resonant mass detectors with a stimulated absorption probability $\gamma_0\Delta t\langle a^\dagger a\rangle = O(1)$ could achieve this. Recently it was shown that the technology for quantized response click detectors, while challenging, is not infeasible~\cite{tobar_detecting_2024,shenderov_stimulated_2024}. These detectors can also operate in the kilo-Hz frequency range of gravitational radiation, with energy densities comparable to those currently observed by LIGO. %It would also allow further statistical null tests using detector cross-correlations between two interferometric detectors~\cite{Maulikcross}.  

A major advantage of the tests we proposed here, based on independent detector cross-correlations, is that they naturally exclude both the vacuum noise of the detectors and excess quantum measurement noise (even with heterodyne detection, which is notorious for adding measurement noise)~\cite{arthurs_simultaneous_1965,loudon_squeezed_1987,clerk_introduction_2010,caves_quantum_1982,caves_quantum_2012,bergeal_phase-preserving_2010,haus_quantum_1962,manikandanFuel,manikandan_complementary_2025}. Used in parallel with statistics of measurements on individual bars proposed in Refs.~\cite {manikandan_complementary_2025,Manikandan_Wilczek_acoherence,Essay}, they provide additional, independent means to probe the failure of the coherent state hypothesis for the gravitational radiation field. Finally, let us recall that detection of any deviation from the coherent state hypothesis tests for gravitational radiation would demonstrate the inadequacy of a classical or semi-classical description, and bring in quantum theory. 

\section{Acknowledgments}
FW is supported by the U.S. Department of Energy under grant Contract Number DE-SC0012567 and by the Swedish Research Council under Contract No. 335-2014-7424. SKM was supported in part by the Swedish Research Council under Contract No. 335-2014-7424 and in part by the Wallenberg Initiative on Networks and Quantum Information (WINQ).  We thank Maulik Parikh and Igor Pikovski for stimulating discussions around these subjects.  
\appendix
\section{Interaction between the radiation field and two identical detectors\label{AppA}}
Consider the interaction Hamiltonian,
\begin{equation}
    H_{I}\Delta t/\hbar =\sqrt{\gamma_0\Delta t}(a^\dagger b_1+b_1^\dagger a+a^\dagger b_2+b_2^\dagger a)=\sqrt{2\gamma_0\Delta t}(a^\dagger d_{+}+d_{+}^\dagger a),
\end{equation}
where $a$ refers to the radiation field, $\gamma_0$ is the spontaneous emission rate of the detector, and $d$ refers to the normal mode of two detectors,
\begin{equation}
  d_{+}=\frac{1}{\sqrt{2}}(b_1+b_2):~[d_{+},d_{+}^\dagger] = 1.  
\end{equation}
We are interested in evaluating,
\begin{equation}
    e^{-iH_{I}\Delta t/\hbar}|\alpha\rangle_{a}|0\rangle_{1}|0\rangle_{2}.
\end{equation}
Note that $d_{\pm}|0\rangle_{1}|0\rangle_{2} = \frac{1}{\sqrt{2}}(b_{1}\pm b_{2})|0\rangle_{1}|0\rangle_{2} = 0,$ meaning that $|0\rangle_{d_{+}}|0\rangle_{d_{-}}=|0\rangle_{1}|0\rangle_{2}$, where we have also defined,
    \begin{equation}
  d_{-}=\frac{1}{\sqrt{2}}(b_1-b_2):~[d_{-},d_{-}^\dagger] = 1.  
\end{equation}
The interaction therefore reduces to,
\begin{eqnarray}
    &&e^{-iH_{I}\Delta t/\hbar}|\alpha\rangle_{a}|0\rangle_{d_{+}}|0\rangle_d{_{-}} =e^{-i\sqrt{2\gamma_0\Delta t}(a^\dagger d_{+}+d_{+}^\dagger a)}|\alpha\rangle_{a}|0\rangle_{d_{+}}|0\rangle_d{_{-}}\nonumber\\
    &=&|\alpha\cos(\sqrt{2\gamma_0\Delta t})\rangle_{a}|-i\alpha\sin(\sqrt{2\gamma_0\Delta t})\rangle_{d_{+}}|0\rangle_{d_{-}}\nonumber\\
    &=&|\alpha\cos(\sqrt{2\gamma_0\Delta t})\rangle_{a}e^{-i\alpha\sin(\sqrt{2\gamma_0\Delta t})(b_1^\dagger+b_2^\dagger)/\sqrt{2}+i\alpha^*\sin(\sqrt{2\gamma_0\Delta t})(b_1+b_2)/\sqrt{2}}|0\rangle_{d_{+}}|0\rangle_{d_{-}}\nonumber\\
    &=&|\alpha\cos(\sqrt{2\gamma_0\Delta t})\rangle_{a}e^{-i\alpha\sin(\sqrt{2\gamma_0\Delta t})(b_1^\dagger+b_2^\dagger)/\sqrt{2}+i\alpha^{*}\sin(\sqrt{2\gamma_0\Delta t})(b_1+b_2)/\sqrt{2}}|0\rangle_{1}|0\rangle_{2}\nonumber\\
    &=&|\alpha\cos(\sqrt{2\gamma_0\Delta t})\rangle_{a}e^{-i\alpha\sin(\sqrt{2\gamma_0\Delta t})b_1^\dagger/\sqrt{2}+i\alpha^{*}\sin(\sqrt{2\gamma_0\Delta t})b_1/\sqrt{2}}|0\rangle_{1}e^{-i\alpha\sin(\sqrt{2\gamma_0\Delta t})b_2^\dagger/\sqrt{2}+i\alpha^{*}\sin(\sqrt{2\gamma_0\Delta t})b_2/\sqrt{2}}|0\rangle_{2}\nonumber\\
    &=&|\alpha\cos(\sqrt{2\gamma_0\Delta t})\rangle_{a}|-i\alpha\sin(\sqrt{2\gamma_0\Delta t})/\sqrt{2}\rangle_{1}|-i\alpha\sin(\sqrt{2\gamma_0\Delta t})/\sqrt{2}\rangle_{2}\nonumber\\
    &\approx&|\alpha\rangle_{a}|-i\alpha\sqrt{\gamma_{0}\Delta t}\rangle_1|-i\alpha\sqrt{\gamma_{0}\Delta t}\rangle_2.
\end{eqnarray}
Now for a genetic state of the radiation field, 
\begin{equation}
    \rho_F = \int d^2\alpha P(\alpha)|\alpha\rangle\langle\alpha|,
\end{equation} 
we find that the joint state $\rho_F\otimes|0\rangle\langle 0|_1\otimes |0\rangle\langle 0|_2$ evolves as,
\begin{eqnarray}
    \rho'&=&\int d^{2}\alpha P(\alpha)|\alpha\cos(\sqrt{2\gamma_0\Delta t})\rangle\langle \alpha\cos(\sqrt{2\gamma_0\Delta t})|\otimes |-i\alpha\sin(\sqrt{2\gamma_0\Delta t})/\sqrt{2}\rangle\langle -i\alpha\sin(\sqrt{2\gamma_0\Delta t})/\sqrt{2}|_1\nonumber\\&\otimes& |-i\alpha\sin(\sqrt{2\gamma_0\Delta t})/\sqrt{2}\rangle\langle -i\alpha\sin(\sqrt{2\gamma_0\Delta t})/\sqrt{2}|_2\nonumber\\&\approx& \int d^{2}\alpha P(\alpha)|\alpha\rangle\langle\alpha|\otimes |-i\alpha\sqrt{\gamma_{0}\Delta t}\rangle\langle -i\alpha\sqrt{\gamma_{0}\Delta t}|_1\otimes |-i\alpha\sqrt{\gamma_{0}\Delta t}\rangle\langle -i\alpha\sqrt{\gamma_{0}\Delta t}|_2.
\end{eqnarray}
We can also achieve the result by considering sequential interactions. Note that the time-evolved state after interacting with the first detector is,
\begin{eqnarray}
   \rho(\Delta t_1)&=& e^{-\frac{i}{\hbar}H_{I,1} \Delta t_1}\rho_F\otimes|0\rangle\langle 0|_1  e^{\frac{i}{\hbar}H_{I,1} \Delta t_1}\nonumber\\&=&\int d^2\alpha P(\alpha)|\alpha \cos(\sqrt{\gamma_0 \Delta t_1})\rangle\langle \alpha \cos(\sqrt{\gamma_0 \Delta t_1})|\nonumber\\&\otimes& |-i\alpha \sin(\sqrt{\gamma_0\Delta t_1})\rangle\langle -i\alpha \sin(\sqrt{\gamma_0\Delta t_1})|_1.
\end{eqnarray}
Now, a second detector interacts with the radiation field in sequence, changing the state to,
\begin{eqnarray}
 \rho' &=& \rho(\Delta t_1+\Delta t_2)= e^{-\frac{i}{\hbar}H_{I,2} \Delta t_2}\rho(\Delta t_1) \otimes|0\rangle\langle0|_2 e^{\frac{i}{\hbar}H_{I,2} \Delta t_2}\nonumber\\&=&\int d^2\alpha P(\alpha)|\alpha \cos(\sqrt{\gamma_0 \Delta t_1})\cos(\sqrt{\gamma_0 \Delta t_2})\rangle\langle \alpha \cos(\sqrt{\gamma_0 \Delta t_1})\cos(\sqrt{\gamma_0 \Delta t_2})|\nonumber\\&\otimes& |-i\alpha \sin(\sqrt{\gamma_0\Delta t_1})\rangle\langle -i\alpha \sin(\sqrt{\gamma_0\Delta t_1})|_1\nonumber\\&\otimes& |-i\alpha\cos(\sqrt{\gamma_0 \Delta t_1}) \sin(\sqrt{\gamma_0\Delta t_2})\rangle\langle -i\alpha \cos(\sqrt{\gamma_0 \Delta t_1})\sin(\sqrt{\gamma_0\Delta t_2})|_2\nonumber\\&\approx& \int d^{2}\alpha P(\alpha)|\alpha\rangle\langle\alpha|\otimes |-i\alpha\sqrt{\gamma_{0}\Delta t}\rangle\langle -i\alpha\sqrt{\gamma_{0}\Delta t}|_1\otimes |-i\alpha\sqrt{\gamma_{0}\Delta t}\rangle\langle -i\alpha\sqrt{\gamma_{0}\Delta t}|_2,
\end{eqnarray}
where in the last step, we have assumed, $\Delta t_1=\Delta t_{2}=\Delta t.$
\bibliography{refe}
\end{document}